\documentclass[fleqn,12pt]{SelfArx} 

\definecolor{color1}{RGB}{0,90,30} 
\definecolor{color2}{RGB}{20,20,0} 

\newlength{\tocsep} 
\usepackage{lipsum} 
\usepackage{natbib}
\usepackage{hyperref}
\usepackage{amssymb,amsmath,amsthm,enumitem}
\usepackage{blindtext}

\usepackage{upgreek}

\renewcommand{\arraystretch}{1.0}

\JournalInfo{Astro2020 APC White Paper: Technology Development Activity} 
\Archive{10 July 2019} 

\PaperTitle{Enabling the next generation of scientific discoveries by embracing photonic technologies} 

\Authors{\normalsize 
Nemanja Jovanovic\textsuperscript{1}**, 
Charles Beichman\textsuperscript{1,2,3}, 
Cullen Blake\textsuperscript{4}
Michael Bottom\textsuperscript{5},
Jeffrey Chilcote\textsuperscript{6},
Carl Coker\textsuperscript{2},
Jonathan Crass\textsuperscript{6}, 
Justin R. Crepp\textsuperscript{6},
Nick Cvetojevic\textsuperscript{7},
Miguel Daal\textsuperscript{8},
Mario Dagenais\textsuperscript{9},
Kristina Davis\textsuperscript{8},
Richard Dekany\textsuperscript{1}, 
Don Figer\textsuperscript{10}, 
Michael P. Fitzgerald\textsuperscript{11}, 
Pradip Gatkine\textsuperscript{9},
Olivier Guyon\textsuperscript{12},
Sam Halverson\textsuperscript{13},
Robert J. Harris\textsuperscript{14},
Philip M. Hinz\textsuperscript{15},
David Hover\textsuperscript{1},
Andrew W. Howard\textsuperscript{1},
Rebecca Jensen-Clem\textsuperscript{15}, 
Jeffrey Jewell\textsuperscript{2},
Colby Jurgenson\textsuperscript{16},
Stephanie Leifer\textsuperscript{2}
Julien Lozi\textsuperscript{12}, 
Stefan Martin\textsuperscript{2},
Frantz Martinache\textsuperscript{7},
Dimitri Mawet\textsuperscript{1,2},
Ben Mazin\textsuperscript{8},
Bertrand Mennesson\textsuperscript{2},
Renan Moreira\textsuperscript{18},
Jacklyn Pezzato\textsuperscript{1}, 
Peter Plavchan\textsuperscript{19}, 
Michael D. Porter\textsuperscript{1},
Garreth Ruane\textsuperscript{2}, 
David Redding\textsuperscript{2}, 
Ananya Sahoo\textsuperscript{12},
Christian Schwab\textsuperscript{20},
Eugene Serabyn\textsuperscript{2}, 
Warren Skidmore\textsuperscript{21},
Andrew Skemer\textsuperscript{15},
David Van Buren\textsuperscript{2}, 
Gautam Vasisht\textsuperscript{2},
Sylvain Veilleux\textsuperscript{9},
Sebastien Vievard\textsuperscript{12},
Jason Wang\textsuperscript{1}
Ji Wang\textsuperscript{16}
}

\affiliation{\textsuperscript{1}\textit{California Institute of Technology}
\textsuperscript{2}\textit{Jet Propulsion Laboratory}
\textsuperscript{3}\textit{NASA Exoplanet Science Center}
\textsuperscript{4}\textit{University of Pennsylvania}
\textsuperscript{5}\textit{University of Hawaii}
\textsuperscript{6}\textit{University of Notre Dame}
\textsuperscript{7}\textit{Observatoire de la Cote d’Azur}
\textsuperscript{8}\textit{University of California Santa Barbara}
\textsuperscript{9}\textit{University of Maryland}
\textsuperscript{10}\textit{Rochester Institute of Technology}
\textsuperscript{11}\textit{University of California Los Angeles}
\textsuperscript{12}\textit{Subaru Telescope}
\textsuperscript{13}\textit{Massachusetts Institute of Technology}
\textsuperscript{14}\textit{Zentrum f\"ur Astronomie der Universit\"at Heidelberg}
\textsuperscript{15}\textit{University of California Santa Cruz}
\textsuperscript{16}\textit{Ohio State University}
\textsuperscript{17}\textit{University of California Berkeley}
\textsuperscript{18}\textit{Ultra-low Loss Technologies}
\textsuperscript{19}\textit{George Mason University}
\textsuperscript{20}\textit{Macquarie University}
\textsuperscript{21}\textit{Thirty Meter Telescope}
}

\affiliation{** \textbf{E-mail, phone}: nem@caltech.edu, +1 (626) 395-1214} 
\newcommand{\keywordname}{Keywords} 

\Abstract{
The fields of Astronomy and Astrophysics are technology limited, where the advent and application of new technologies to astronomy usher in a flood of discoveries altering our understanding of the Universe (e.g., recent cases include LIGO and the GRAVITY instrument at the VLTI). Currently, the field of astronomical spectroscopy is rapidly approaching an impasse: the size and cost of instruments, especially multi-object and integral field spectrographs for extremely large telescopes (ELTs), are pushing the limits of what is feasible, requiring optical components at the very edge of achievable size and performance. For these reasons, astronomers are increasingly looking for innovative solutions like photonic technologies that promote instrument miniaturization and simplification, while providing superior performance. 
\smallskip

Astronomers have long been aware of the potential of photonic technologies. Despite this, the field has been slow to adopt photonics solutions for major facility instruments. There are four possible explanations:1) Difficulty of efficient optical coupling, 2) Cost and effort of re-engineering existing industrial photonics for niche astronomy applications, 3) Lack of technological maturity of certain photonic components and technologies, and 4) A conservative mentality amongst astronomical instrumentation designers, instrument review panels, and funding bodies.
\smallskip

The goal of this white paper is to draw attention to key photonic technologies and developments over the past two decades and demonstrate there is new momentum in this arena. We outline where the most critical efforts should be focused over the coming decade in order to move towards realizing a fully photonic instrument. A relatively small investment in this technology will advance astronomical photonics to a level where it can reliably be used to solve challenging instrument design limitations. For the benefit of both ground and space borne instruments alike, an endorsement from the National Academy of Sciences decadal survey will ensure that such solutions are set on a path to their full scientific exploitation, which may one day address a broad range of science cases outlined in the KSPs.}
\begin{document}

\flushbottom 

\maketitle 



\clearpage


\section{An Introduction to Photonics}
\smallskip
In order to understand the potential of photonic technologies, it is important to first understand what photonics are and the properties that make them favorable for astronomy. There is no single definition of what falls under the category of photonics, so we adopt the following commonly accepted definition here: \textbf{Photonics are the class of components and devices that either makes use of optical waveguides (optical fibers or integrated circuits) or micro/nanostructured materials (e.g. photonic crystals or metamaterials) that can manipulate the optical properties of light (phase, direction, etc.).}

In this paper, we confine our discussion to waveguiding photonic technologies. Waveguiding photonic components such as optical fibers and optical integrated circuits have been advanced over the past 30 years by significant investment by the telecommunications industry. This has led to the maturation of many technologies, with components now being widely used in many other fields, including astronomy. However, the overwhelming majority of these components were optimized for operations in the telecommunications O- and C-band wavelength ranges (1260--1360~nm and 1530--1565~nm) which does not address the needs of the broad astronomical community. 

Light can be routed with a photonic waveguide (fiber or chip platform) thanks to the principle of total internal reflection (amongst others). This requires that the core region, where the light is to be guided, have a higher refractive index than the surrounding material, known as the cladding. Depending on the core size, the index contrast between the core/cladding, and core/
cladding structure, the waveguide will fall into one of two categories: multimode (MM) or single-mode (SM). SM devices are named as such because they can only transmit light in a single spatial mode, which is typically categorized by a near-Gaussian profile and a flat phase front. These devices have core sizes typically between 100 nm and $20~\upmu$m in size and operate at the diffraction limit. In contrast, MM devices support 100s-1000s of modes with distinct irradiance and phase profiles in cores that can be as large as $1$~mm in diameter. 

\subsection{Coupling to Photonic Waveguides}\label{sec:coupling}
Given their small spatial extents, efficiently coupling into waveguides is a key consideration when deciding their applicability for astronomical applications. The coupling efficiency depends on the extent to with which one can match the input beam properties (irradiance and phase) to those of the modes supported by the waveguide. The plethora of modes supported by a MM device allows even complex beam profiles to be efficiently coupled, which is why they have been referred to as a ``bucket for light" and been the fiber technology of choice for seeing-limited spectrographs. 

On the other hand, matching the beam from a telescope to the requirements of a SM device is challenging. In the presence of atmospheric turbulence, the flat phase front requirement of the mode can only be met for ground-based observations through the use of an adaptive optics (AO) system. Significant effort has been invested in AO fiber injection over the past 20 years~\citep{coude2000,Perrin194,mennesson2010,ghas2012,bechter2016}, but despite many successes, delivered Strehl ratios of $30-50\%$ currently available at most telescopes limit SM fiber coupling efficiencies. More recently, several large observatories have commissioned ``extreme" AO (ExAO) systems capable of delivering $90\%$ Strehl ratios in better-than-median seeing~\citep{dekany2013,Macintosh2014,jovanovic2015,vigan2016}, well-suited for photonics applications.  

High Strehl ratio, however, still leaves a beam shape mismatch between the Airy beam from the telescope and the Gaussian-like mode of a SM fiber. The theoretical coupling limit of $\sim80\%$~\citep{Shaklan1988} is reduced by the introduction of a central obstruction, and continues to drop as the central obstruction size increases. To prevent this beam mismatch, the pupil can be apodized with optics in a lossless fashion to match the illumination profile of the fiber. This was recently demonstrated by~\cite{jovanovic2017a}, who boosted the theoretical coupling efficiency to $>90\%$ using this technique. Indeed,~\cite{jovanovic2017a} were the first to demonstrate the simultaneous achievement of both requirements outlined above using the SCExAO instrument on the Subaru Telescope~\citep{jovanovic2015} and the subsequent realization of $>50\%$ coupling efficiencies to SM fibers from large apertures on the ground. This milestone demonstration has validated that it is indeed possible to achieve the challenging combination of coupling light into SM fibers from large-aperture ground-based telescopes. Although it relies on sophisticated AO systems, this is increasingly common at all large observatories. With high coupling efficiencies for both MM and SM fibers clearly established, either family of photonic waveguides can now be considered for collection and routing of light to an instrument.

\subsection{Advantages of SM Waveguides}\label{sec:smwaveguides}
Owing to the single spatial mode supported by a SM waveguide, the output of such a device does not evolve with time or with input beam illumination fluctuations, and so is extremely stable. This effect is termed ``spatial filtering" and is analogous to passing a beam through a very small pinhole. SM fiber use offers a clear advantage for instruments that require a stable instrument point-spread function (PSF). For example, it would eliminate modal noise as well as pointing-induced illumination fluctuations common to MM fiber-fed radial velocity (RV) spectrographs. A stable instrument response function makes it possible to calibrate data to a greater precision improving SNR for a given observation. It also means that the performance of the instrument is entirely decoupled from the performance of the telescope, AO system, and injection, which can greatly relax constraints on the design of the telescope and upstream optics. This decoupling reduces both cost and complexity, powerful motivators that have led future space missions such as EarthFinder~\citep{plavchan2018} to plan for use of SM fibers to detect and study Earth-analogs in the habitable zone around G-stars from space. 

Operating at the diffraction-limit also offers critical benefits to the spectrograph design and ultimately its size~\citep{jovanovic2016}. To understand this, consider the basic equations governing spectrograph performance. Equations~\ref{equ:resolution} and~\ref{equ:R} show the resolution ($\delta\lambda$) and resolving power ($R$) of a spectrograph, respectively~\citep{schroeder1999}. 
\setlength{\belowdisplayskip}{3pt}
\setlength{\belowdisplayshortskip}{3pt}
\setlength{\abovedisplayskip}{3pt}
\setlength{\abovedisplayshortskip}{3pt}
\begin{equation}
    \delta\lambda = \frac{r\phi D_{\rm{Tel.}}}{Ad_{\rm{col.}}}
    \label{equ:resolution}
\end{equation}
\begin{equation}
    R=\frac{\lambda}{\delta\lambda} = \frac{\lambda Ad_{\rm{col.}}}{r\phi D_{\rm{Tel.}}}
    \label{equ:R}
\end{equation}
The parameters in the equations are defined as follows: $\lambda$ is the wavelength of light, $r$ is the anamorphic factor, $\phi$ is the angular slit size, $D_{\rm{Tel.}}$ is the diameter of the telescope, $A$ is the angular dispersion, and $d_{\rm{col.}}$ is the diameter of the collimated beam incident on the disperser. It can be seen that the resolving power is inversely dependent on the slit size and is maximized when the slit size is at a minimum (i.e. the diffraction-limit). It can also be seen that the resolving power will decrease as the telescope diameter increases. For a given wavelength, angular dispersion, slit size, and anamorphic factor, the diameter of the collimator (and hence disperser and camera optic) needs to proportionally increase to maintain a certain resolving power. To preserve the focal ratio of the collimator and the magnification of the spectrograph, the focal length of both the collimator and camera optics must also increase proportionally. This ultimately drives an increase in the volume of the entire seeing-limited spectrograph. However, SMFs operate at the diffraction-limit, so the slit formed by the output beam of such a fiber would have a size that can be expressed as
\begin{equation}
    \phi \approx \frac{\lambda}{D_{\rm{Tel.}}}.
    \label{equ:diffractionlimit}
\end{equation}
Substituting Equation~\ref{equ:diffractionlimit} into Equations~\ref{equ:resolution} and~\ref{equ:R}, we obtain
\begin{equation}
    \delta\lambda = \frac{r\lambda}{Ad_{\rm{col.}}}
    \label{equ:resolution2}
\end{equation}
and 
\begin{equation}
    R=\frac{\lambda}{\delta\lambda} = \frac{Ad_{\rm{col.}}}{r}.
    \label{equ:R2}
\end{equation}

In the diffraction-limited regime, it is clear that both parameters are now independent of the telescope diameter and are only a function of the optics in the spectrograph itself (i.e., the angular dispersion and collimated beam size). This means that a high-resolution spectrograph which exploits a diffraction-limited feed (e.g. a SMF) would have the same properties independent of the observatory, and the instrument can be designed without consideration of the telescope. \textit{\textbf{It also means that that the volume of the spectrograph no longer scales with telescope aperture for a given resolving power, and very high resolution spectrographs can be made extremely compact.}} Furthermore, thermal and vibration deflections, which scale with the cube of length of the instrument, are also significantly reduced, providing exceptional instrument stability. These advantages were first exploited by the RHEA~\citep{feger2016} and iLocator~\citep{crepp2016} instruments and more recently the upcoming HISPEC/MODHIS spectrographs. HISPEC/MODHIS is a SMF-fed spectrograph operating from y-K band designed for operation at $R>100,000$ at both the Keck and TMT Observatories. By operating at the diffraction limit this instrument will cost several times less than other first-light seeing-limited instruments under development for ELT's.

\subsection{Other Photonics Properties}\label{sec:photproperties}
In addition to spatial filtering, photonics technologies support numerous other functions typically seen with classical bulk optics such as spectral filtering, spectral dispersion, frequency comb generation, beam combination and even interferometric starlight nulling. When you combine these properties with the compact footprint of photonic devices (optical fibers measure $250~\upmu$m in diameter and photonic chips of only a few square centimeters) and often monolithic (single component) platforms, waveguiding photonic devices offer significant potential for the miniaturization of astronomical instrumentation.

In this paper we offer an overview of several key areas in astronomy photonics have completely transformed, outline others where significant development has been undertaken and what the critical next steps are. The goal is to highlight areas where development should be focused over the next decade to enable the possibility for fully photonic instruments in the future. These instruments offer the opportunity for entirely new and transformative science. A complementary whitepaper \citep{affordable2019} makes the case that photonics applied at the observatory level may offer a cost-effective approach to ever larger systems.

\section{Areas Impacted by Photonics}
\smallskip
Here we describe two key areas that have most widely embraced and benefited from the application of photonics over the past two decades. While we focus on facility-level instrumentation in this section, it would be amiss not to mention the substantial development of astrophotonic technology currently at the prototype/visitor instrument level, examples of which can be found in a recent special issue of Optics Express~\citep{AstroPhotSpecialIssue2018}.

\textbf{Large scale surveys enabled by MMF fed spectrographs:} There are numerous advantages to feeding a spectrograph with an optical fiber, including: 1) the ability to locate the spectrograph in a more stable location without the need for complex beam steering optics, 2) the ability to maximize the use of detector real estate in a multi-object spectrograph by aligning the output fibers vertically along a pseudo-slit, leading to much higher efficiency 3) the ability to feed light into a vacuum vessel without needing a window and 4) the ability to control the input illumination to the spectrograph due to the fixed fiber positions along the slit or scrambling/homogenization. These advantages were recognized early on, and in the late 1970s, the first fiber fed spectrographs were demonstrated~\citep{hill1980,hill1988}. This technical development promoted countless new instruments with advanced capabilities over the next 20 years. For example, by combining fibers with a robotic fiber positioner, the 2dF Galaxy Redshift Survey team increased the observational efficiency of the AAOmega instrument by several orders of magnitude, enabling them to produce one of the first high-redshift maps of galaxies in the Universe~\citep{colless2001,croom2004} and paving the way for an era of very large-scale surveys ($100,000+$ objects). 


\textbf{Precision astrometry with interferometers:} Interferometry offers the possibility to combine light from multiple telescopes to increase the angular resolution beyond what is imposed by the size of a single aperture. Once the light from the telescopes is routed to a central room it is then combined to form interference fringes which can be analyzed to study the astrophysical object. 

\begin{figure*}[h]
    \centering
	\includegraphics[width=0.99\textwidth]{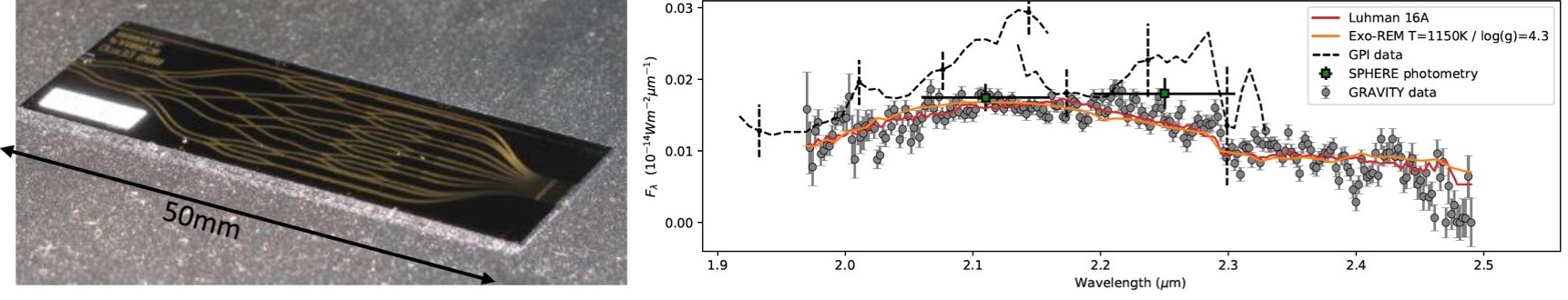}
	\centering
	\caption{\footnotesize{(Left) An image of the integrated photonic beam combiner used by GRAVITY. The circuit which is used to combine the light from the four 8-m telescopes at the VLT via a series of splitters and couplers can clearly be seen. The wafer is approximately the size of two US quarters and monolithic (i.e. a single component) as opposed to classical bulk optic beam combiners.  This image was reproduced based on Figure 3 in~\cite{perraut2018}. This device was designed and manufactured by CEA/LETI (Grenoble, France).  (Right) the superior spectrum obtained for HR8799 e with GRAVITY compared to previous results. This figure was reproduced from the~\cite{lacour2019}.}}
	\label{fig:beamcomb}
\end{figure*}

It was recognized early on that the spatial filtering property of a SM waveguide/fiber could be used to improve the contrast of the generated fringes and hence maximize the SNR of the data \citep{coude1992,coude1994}. The European Southern Observatory (ESO) invested in many generations of photonic beam combiners (both fiber- and chip-based platforms including MIDI~\citep{MIDI2003}, AMBER \citep{AMBER2007}, and PIONIER~\citep{PIONIER2011}) culminating with the GRAVITY instrument~\citep{gravity2017}. GRAVITY utilizes a 4-channel integrated photonic beam combining chip to combine the light of the 8-m telescopes at the VLT~\citep{perraut2018} (see Figure~\ref{fig:beamcomb} for details). 
The long baselines and exquisite calibration of the interferometer enable micro-arcsecond precision astrometry. This extreme performance has been truly tranformative. The instrument has greatly improved the astrometry on the stars orbiting the supermassive black hole at the center of the Milky Way and have enabled the first observation of flares from the accretion disk at the edge of the event horizon~\citep{gravity2018flares}. In the exoplanets field, GRAVITY has delivered a high resolution spectrum of HR8799e and the most precise astrometry for any imaged exoplanet to date~\citep{lacour2019}.

\section{The Road to Photonic Spectroscopy - State-of-the-field}
\smallskip
Despite the advances from photonic devices in the areas of beam combination, nulling, and interferometry, the most promising underexploited avenue for astrophotonics is spectroscopy. While certain photonic technologies are already being applied to spectroscopy, in particular the use of optical fibers, the transformational potential of fully photonic spectrographs operating at the diffraction-limit, enabled by their size, simplicity, and low unit cost has yet to be realized.


\subsection{The Photonic Spectrograph:} 
\begin{figure}[ht!]
	\includegraphics[width=0.49\textwidth]{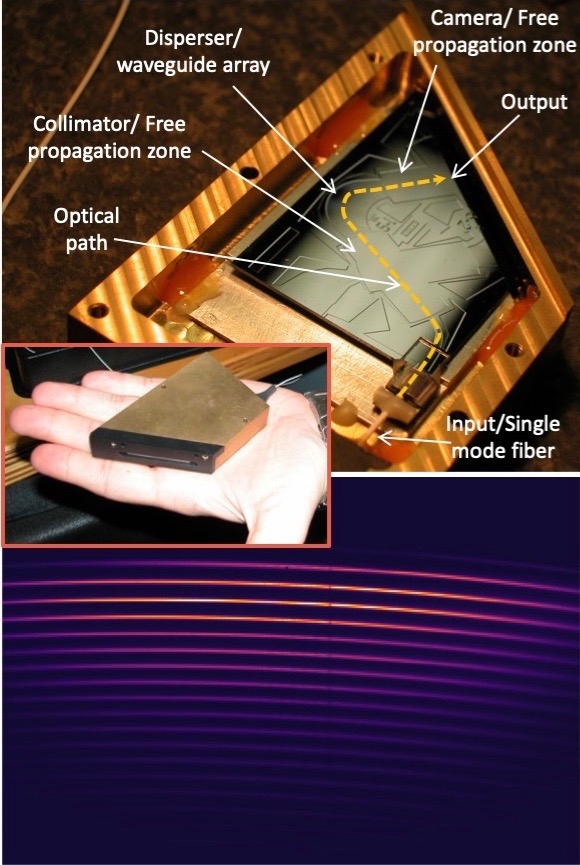}
	\caption{\footnotesize{(Top) An arrayed waveguide grating version of an integrated photonic spectrograph device. An off-the-shelf device with a $R=7000$, packaged and connected to a SMF feed. The circuit on the chip is clearly visible~\citep{cvetojevic2009}. (Inset) An image for sense of scale. Disclaimer, to the best of the teams knowledge the hand used in this image is of average size. (Bottom) A J and H band spectrum of a lamp taken with the AWG prototype instrument in the laboratory~\citep{jovanovic2017d}.}} 
    \label{fig:Photonicspectrograph}
\end{figure}

\smallskip
An Integrated Photonic Spectrograph (IPS) is a compact device, typically measuring a few square centimeters, which consists of an optical circuit imprinted into a transmissive material on a wafer. An example of a device with an $R$ of $\sim7000$ operating around $1.55~\upmu$m is shown in Figure~\ref{fig:Photonicspectrograph}. Owing to the fact they are fed with a diffraction-limited SMF, these devices include the equivalent of a spectrograph  collimator, disperser, and camera optics in a single monolithic element, which fits in the palm of your hand, enabling an order of magnitude reduction in each of the $3$ linear instrument dimensions (3 orders of magnitude smaller volume) with respect to a bulk-optic counterpart.   

Several IPS variants exist including Planar Concave Gratings and Arrayed Waveguide Gratings (AWGs) to name a few (please see~\cite{gatkine2019} for a review on the topic). Given that AWGs are the most mature versions of IPSs, development and testing for astronomical applications has almost exclusively been reserved to them. They were originally developed by the telecommunications industry and are now used in optical networks worldwide. Off-the-shelf devices like the one shown in Figure ~\ref{fig:Photonicspectrograph} have been optimized to operate around $1550$~nm, with high throughput ($78\%$), a free-spectral range (FSR) of $50$~nm (defined as the wavelength span of a single order of the dispersion element), and medium $R\sim7000$~\citep{cvetojevic2012a}. More recently there have been efforts to extend the operating range of such devices for biomedical applications~\citep{geuz2017}. This milestone demonstration successfully realized an AWG that could operate from $400$--$1700$~nm in a single shot at low resolving power ($R\sim100$).

Over the past 10 years, AWG-based IPS technologies have been tested for astronomical applications. The off-the-shelf device mentioned above was tested in several prototypical instruments on-sky. \cite{cvetojevic2009} first used the device to generate a full H-band spectrum of night time airglow by pointing the collection SMF towards the sky (no optics or telescope) and using a bulk optic cross-disperser to unravel the orders. The team showed that it was possible to use an AWG to process signals from multiple SMFs simultaneously in a single chip and recover them through cross-dispersion~\citep{cvetojevic2012a}. This concept was tested at the Australian Astronomical Telescope (AAT)~\citep{cvetojevic2012b}. 
To take full advantage of the technology, the highly efficient SMF injection built behind SCExAO (outlined above) was used. With $>50\%$ coupling efficiency from an 8-m telescope on the ground directly into a SMF realized~\citep{jovanovic2017a}, it was possible to take the ultimate step and directly feed the IPS device. The team used the spectrograph once again with external cross-dispersion to enable operation across the J and H-bands with a total throughput from fiber to detector of $42\pm3\%$ and a resolving power of $\sim5000$~\citep[$5\%$ throughput from sky-to-detector assuming a 65\% Strehl ratio;][]{jovanovic2017d}. The photonic instrument successfully recovered spectra from several stellar sources. This  demonstration shows that it is now possible to consider a fully photonic instrument that can be competitive with traditional approaches. 

\textit{\textbf{Although this initial success is encouraging, the first demonstration device lacked the specifications suited to most astronomical applications. Having proven the fundamental advantages, we need to further the development into engineered components for astrophysics.}} NASA appreciates the importance of this development and understands this technology could revolutionize space missions and awarded an APRA to support some ongoing efforts in this domain~\citep{Gatkine2017}, but significantly investment is needed.

\subsection{Enabling Photonic Spectroscopy:} 
\smallskip
The spectrograph is the single element that if replaced with a photonic alternative could dramatically alter the course of astronomical instrumentation. There are numerous photonic technologies which can be used to either couple the light from a telescope that does not have an ExAO system, realize an IFU that can feed an IPS, spectrally filter the light or enhance calibration. The next two sections summarize some of the devices that enable and enhance an IPS.

\textbf{Photonic lanterns:} these devices allow light from a low Strehl ratio beam to be coupled efficiently at the focus of the telescope while providing a diffraction limited input to a downstream spectrograph~\citep{sergio2013,birks2015}. Originally, they were produced by tapering a bundle of SMFs inserted into a glass capillary (melting and drawing) until they combined to form a single MMF which could efficiently couple the low Strehl beam at the focal plane. As long as the number of SMFs is equal to or exceeds the number of modes supported in the MMF end, there will be an efficient transition from a seeing limited beam to many diffraction limited beams offering diffraction-limited performance to a spectrograph on a poor AO/seeing-limited telescope (these devices can of course also be used on high Strehl ratio beams, but at high Strehl ratio there are advantages of going directly with SMFs). 

Photonic lantern devices based on fibers and integrated photonic chips have been demonstrated \citep{sergio2005,thomson2011,birks2012} and optimized for high efficiency in the H-band~\citep{Noordegraaf2009,Noordegraaf2012,jovanovic2012,Spaleniak2013}. More recently, both types of devices were tested on-sky on seeing limited~\citep{trinh2013} and AO equipped telescopes~\citep{maclachlan2014,harris2015,Cvetojevic2017,jovanovic2017a}. The integrated photonic demonstrations also reformatted the output of the device to a linear slit of SM waveguides, in preparation to feed a spectrograph. 



One interesting development in photonic lantern technology has been the invention of the mode selective lantern~\citep{sergio2014}. This device allows for one particular spatial mode at the MM end of the device to couple to only one SMF at the output. By addressing spatial modes in this way it may be possible to exploit these devices for imaging or wavefront control as well.


\textbf{Multicore fibers:} we define a multi-core fiber (MCF) as any fiber, which supports numerous cores, be it SM of MM within a single cladding and is formed with tapering/fusing techniques (to distinguish from stacking fibers in a tube and gluing them in place) (this follows the discussion in~\cite{Jovanovic2017c}. Astronomers recognized early on that a fiber with multiple cores could be used to transfer spatial information from the focal plane to the spectrograph, just like an integral field unit (IFU). 



A recent application involves using SMF based MCFs as IFUs to study  directly imaged exoplanets. An AO system and a coronagraph are first used to suppress the bulk of the starlight, which is many orders of magnitude brighter than the planet. Then a SMF (or one core of a MCF) is placed at the location of the known planet~\citep{snellen2015}. Due to the strict requirements of coupling into a SMF, the unwanted starlight coupling is reduced by another factor of 2-3~\citep{mawet2017}. The signal can then be dispersed in a spectograph and integrated upon. The other fibers in the input focal plane can be used to track the stellar flux and collect a spectrum of the star with telluric contamination contemporaneously, which can be used in the spectral extraction process~\citep{wang2017}. This method is currently being explored by the SCExAO~\citep{jovanovic2017b} and KPIC instruments~\citep{mawet2018}. Recent simulations by~\cite{coker2019} have shown that SMFs have the potential to expand the usable spectral bandwidth over which high contrast is maintained by a factor of 2-3 when used in this manner. Another recently proposed instrument concept combines a vector Apodized Phase Plate (vAPP) with a micro-lens array and a MCF to realize the SCAR coronagraph~\citep{por2018,haffert2018}. 



\textbf{Endlessly single-mode fibers:} these optical fibers defeat the finite bandwidth limitations of standard SMFs and allow a fiber to perform as a SMF across the entire range of transmission and could find future applications~\citep{Birks1997}. 


\subsection{Spectral Filtering \& Calibration}
Spectral filtering can be divided into two main categories: filtering science light or filtering a calibration lamp to create reference lines. In addition, reference lines can also be created by a laser.

\textbf{Filtering the science light}: With the light collected by a SMF and routed to a spectrograph it is possible to exploit Fiber Bragg Gratings (FBGs) to conduct spectral filtering of the light prior to injecting it into the spectrograph. They are versatile and can be tailored to remove light from parts of a spectrum with custom bandwidths and attenuation's. These devices were explored for the removal of OH-lines which are caused by the recombination of molecules in the atmosphere at night. The FBGs fabricated for this application were among the most complex spectral filters ever realized and attenuated over 100 lines across the H-band, with varying bandwidths and levels of attenuation~\citep{Bland-Hawthorn2004,Bland-Hawthorn2011,trinh2013}. These FBGs were utilized in the GNOSIS instrument to eliminate the OH line contamination in the spectrum of redshifted galaxies prior to injection into a spectrograph~\citep{Bland-Hawthorn2011}. 


Bragg gratings have also been integrated onto monolithic chips with photonic lanterns to open up the possibility for compactness and scaling in future~\citep{spaleniak2014}.

\textbf{Filtering the calibration light}: Photonic technologies can also be used to filter the spectrum of a broadband lamp to create a comb of lines used for spectral calibration. These devices do not have the stability and accuracy of laser frequency combs but offer a simple, compact and cost effective alternative.

These technologies revolve around using SMF-based interferometers to create a comb-like spectrum for calibration. One version uses two simple fiber splitters/couplers to form a Mach–Zehnder interferometer~\citep{feger2014}. Another is based on a Fabry–Perot etalon formed by a very short length of SMF by using mirror coatings on the ends~\citep{halverson2014}. An integrated-optic version could use micro ring-resonators~\citep{ellis2012}. They all deliver a comb of lines at the output when a broadband light source is injected into the input, the spacing of which can be optimized by modifying the physical dimensions of the devices. As demonstrated by~\cite{gurevich2014} and~\cite{schwab2015}, the etalon version can be locked to an atomic transition as a wavelength reference, resulting in long term stability at the cm/s level from a very simple system. 


\textbf{Laser frequency combs (LFCs)} In 2005 the Nobel Prize in physics was awarded for contributions to the development of laser-based precision spectroscopy, including the optical frequency comb technique. The LFC, a photonic device using nonlinear fibers or chip-based micro-resonators, produces a spectrum consisting of laser lines spaced regularly in frequency that can be employed for spectral calibration. Once phase locked and stabilized, the wavelength of each comb line is stable to better than one part in 10$^{12}$, for years or decades. LFCs address several key limitations in spectral calibration sources: they offer comb lines across the entire spectrum of interest, many more lines for calibration than spectral lamps, and provide more uniformity in the brightness of the lines. 
The first LFCs were validated on instruments like HARPS and demonstrated distinct improvements compared to gas lamp calibration sources. 

New approaches to frequency comb generation have been developed that offer performance more optimally suited (spectral and temporal stability, wavelength range of operation and inter-line spacing) to precision RV (PRV) studies~\citep{yi2016,Obrzud2018,metcalf2019}, but further development is required. Considerable efforts are now being invested to develop compact, low power, chip-based combs suitable for astronomical applications ~\citep{obrzud2019,suh2019}.

Both families of frequency combs described here could be fed into an IPS device contemporaneously with the science signal (in a second SMF) allowing for accurate calibration of the IPS data.

\section{Technology Drivers}
\smallskip
The biggest roadblock to building compact, fully photonic, instruments is being able to customize IPS's for astronomical applications. \textit{\textbf{Given that spectrographs are becoming prohibitively expensive on ELTs, replacing a classical bulk optic spectrograph with an IPS is where the most significant gains can be made in reducing footprint, complexity, and cost and hence where the greatest community energy should be focused in the coming decade}}.

\subsection{Developing IPS Technologies}
\smallskip
Given AWG technology is well understood and has been tested and characterized in the context of astronomy for the past 10 years, it is the ideal choice for the next stage of optimization (although other technologies should be explored in parallel). AWGs should be developed beyond the requirements for the telecommunications industry to support the wide ranging scope of astronomical applications, by engineering them for higher spectral resolution ($R>7000$) than is currently available, in entirely new wavebands spanning from $500$~nm in the visible to $5.0~\upmu$m in the MIR for use in both single and multi-object science cases with on-chip cross-dispersion implemented.  

\begin{figure}
    \centering
	\includegraphics[width=0.49\textwidth]{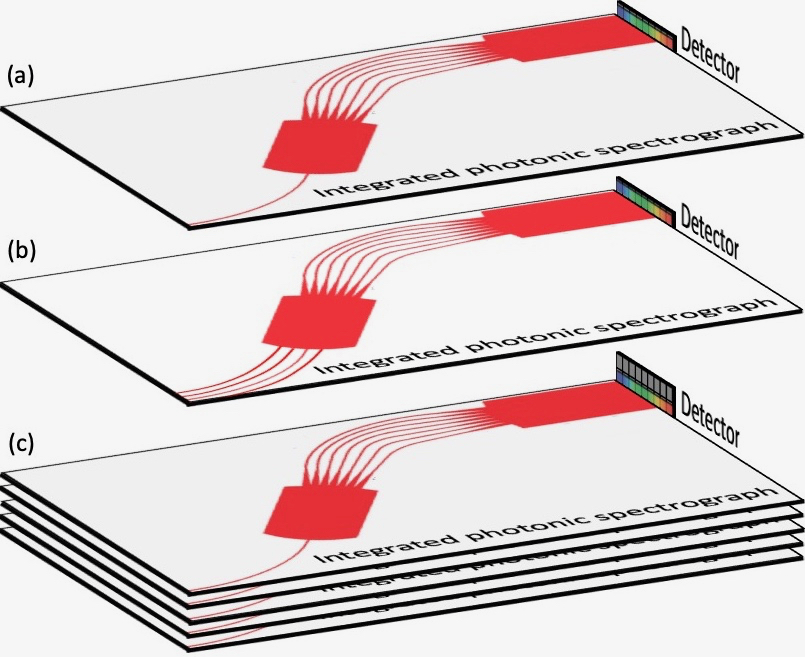}
	\centering
	\caption{\footnotesize{Concept of an AWG instrument (a) AWG butted against a linear detector, (b) AWG with multiple input fibers butted against a linear detector and (d) multiple AWGs butted against a 2D detector~\citep{doug2018}.}}
	\label{fig:widebandAWG}
\end{figure}

By developing custom devices a range of innovative instrument architectures can be considered. The most streamlined concept involves butting the output facet of an AWG directly against the detector, eliminating the need for reimaging optics. Since the output is formatted in a 1D line, a linear array detector could be used reducing costs and complexity (Figure~\ref{fig:widebandAWG}(a)). If the detector were an MKID array~\citep{Szypryt2017}, its ability to resolve the energy of the incoming photon would eliminate the need for cross-dispersion on the chip (a recent NASA STTR was awarded for this development)~\citep{cropper2003}.
To increase the number of objects that can be observed simultaneously several AWGs could be stacked on top of one another utilizing a 2D detector array instead (Figure~\ref{fig:widebandAWG}(c)). This assumes that only a single FSR from each device needs to be imaged by the detector. An alternative method to increase the number of objects that can be studied is to feed the AWGs with multiple input SMFs (see Figure~\ref{fig:widebandAWG}(b)). To unravel the signals from different sources or increase the bandwidth for a single object beyond a single FSR, the orders need to be separated with a cross-disperser which has recently been shown on-chip as well. Here AWGs are cascaded so that the first device splits the light into coarse spectral channels and then a series of high dispersion devices operate on narrower wavelength ranges downstream. To realize these minimal instrument concepts however (i.e. butting chips to detectors), the optical properties of the devices need to first be investigated at the cryogenic temperatures that detectors operate at.

\begin{table*}[ht!]
\caption{\footnotesize{A list of photonic developments that we would like to see over the next 5-10 years to enable photonic spectroscopy. Most devices have thus far been developed in the telecommunications waveband. owing to laboratory equipment availability. and will need to be optimized as part of this process for other wavebands.}}
\begin{tabular}{p{0.15\textwidth}p{0.78\textwidth}}
\hline
\textbf{Technology}         & \textbf{Development description} \\
\hline
 
 IPSs & \footnotesize{IPSs are available from many sources but do not meet astronomical requirements. IPSs should be developed to have higher spectral resolution ($R>7000$) in new wavebands ($500$--$5000$~nm); able to support single/multi-object science cases, on-chip cross-dispersion, and concurrent calibration sources; cryogenic testing of IPS-detector interface.}\\
 
 Photonic lanterns  & \footnotesize{Custom lanterns are only available through a single vendos. Ideally we would increase this vendors ability so they become standard products. Develop new devices with high efficiency across various wavelength ranges (vis-MIR). Test mode selective lanterns for imaging and wavefront control purposes.}    \\
 
 
 Multicore fibers   & \footnotesize{These devices are available through a few vendors optimized for the telecommunications band only. These devices should be developed for other wavelength ranges (vis-MIR) and tested to see if they are suitable to be used for SMF IFUs.}     \\
 
 Endlessly single mode fibers & \footnotesize{Currently only available through a single vendor. These should be developed for application in the MIR, with simulations and experiments to determine their applicability for feeding spectrographs and/or in IFUs with micro-lenses.}    \\
 
 Frequency combs    & \footnotesize{Compact electro-optic and micro-combs are not commercially available. These should be further developed and optimized to include a broader wavelength coverage in the NIR from y-K, devices that even span into the MIR, and a higher uniformity in brightness of the comb lines across the spectrum. }   \\
 
 
 Beam combiners      & \footnotesize{There are many commercial vendors for beam combiner chips. Beam combining chips if correctly designed could be used for nulling. Further development is needed into new waveband (vis to MIR), active phase delays to scan path lengths achromatically, and methods to calibrate these devices.}    \\
 \hline
 \label{tab:developments}
\end{tabular}
\end{table*}

\textbf{Organizations:} There are many commercial foundries that fabricate photonic devices regularly that typically have better control of their process parameters (although University-based foundries can equally play a key role). These foundries offer Photonic Design Kits (PDKs), software to design standard devices compliant with their processes. In addition there are numerous commercially available photonic design suites out there for more customized component design and analysis. To develop astronomically relevant devices, researchers at Universities pursuing this area of research such as Caltech, JPL, RIT, and U Maryland, to name a few, will need to work closely with these foundries to understand their processes and limitations and use the design tools mentioned. A careful choice of material platform will play into this but will most likely involve silica-on-silicon, silicon nitride, and/or silicon-on-insulator. In 10 years time we hope to have strong relationships and a well-developed community procurement process, with several world class fabrication foundries capable of delivering IPSs on demand. 

\textbf{Schedule:} The schedule for IPS development depends on the specific science application. Applications that are similar to current performance specifications can be realized in a much shorter time frame. More ambitious specifications, $R>100, 000$ over a wide bandwidth for example will take time as they are a departure from the current performance capabilities. With a focused development on simple devices, reasonably performing prototype devices could be produced within a 2 to 3-year period (the scope of an NSF or NASA proposal). We expect it to take 5-10 years to build a comprehensive design suite capable of designing custom modifications to IPSs, realizing devices with more challenging properties and understanding limitations of the technology.

\textbf{Cost estimate:} \textit{\textbf{An investment of the order of \$10M in IPS technology over the next 5-10 years would significantly advance the field and make it possible to select IPS approaches for future instruments for both ELTs and space missions}}. This estimate for the IPS development is based on projections from recent costing exercises performed in early 2019 for NASA APRA and JPL RTD grants. It should be highlighted that this development cost pales in comparison to current cost estimate for a single ELT or space mission spectrograph, but is currently considered too risky for funded projects to embrace.  Given the potential for cost reduction and performance improvement, such a public/private level of investment would be a small price to pay.

\subsection{Developing Supporting Technologies}
In this section we briefly describe where we would like to see further development of photonic devices that can enable and enhance photonic spectroscopy over the next 5-10 years. Our findings are summarized in Table~\ref{tab:developments}.

\textbf{Organization:} These technologies need to be developed with a combination of University researchers and vendors. Ultimately we want the technology to transfer to commercial vendors who will conduct the necessary engineering and be able to deliver much more cost effective devices with shorter turn around times in the future. 

\textbf{Schedule:} The devices listed in the table above should be developed steadily over the next 5-10 years consistent with available funding. 

\textbf{Cost estimate:} To engage with industry and get them to develop some of these components commercially will cost more than if done within research laboratories. We estimate that an investment of \$10M would be needed to advance all elements in the table above with commercial vendors (not including the IPSs discussed separately above) to the point where at least the feasibility of each development would be understood. Depending on the difficulties with developing some of these components in new wavelength ranges, which may require new materials, process and equipment, further investments may be needed to fully complete the list.

\section{Cost Estimates}
\smallskip
Combining the cost estimates outlined for technology development, photonic technologies could be significantly advanced with $\leq$\$20~M. (small ground based funding category). More advanced components will naturally get funding support flowing down through instrument trade and risk mitigation studies, but less mature technologies will require directed funding to address key issues before the technology can be considered for such studies. At a fraction of the cost of an instrument for the ELTs or a space mission this investment will pay for itself rapidly.

Despite the fact that GRAVITY at the VLTI requires four 8-m class telescopes, four high performance AO systems, extreme laser calibration metrology, and one of a kind integrated photonic beam combiners, \textit{\textbf{ESO guided by European scientists recognized the potential of their courageous, sustained investment and now have an globally unrivaled science capability.  A similar US commitment to investment in integrated astrophotonics is needed over the next decade to reestablish US astronomy competitiveness in these forefront observational capabilities.}}

\clearpage

\bibliographystyle{apalike}
{\footnotesize
\bibliography{wp.bib}}



\end{document}